\documentclass[10pt,conference]{IEEEtran}
\usepackage[T1]{fontenc}
\usepackage{newtxtext,newtxmath}
\usepackage{graphicx}
\usepackage{booktabs}
\usepackage{amsmath}
\usepackage{url}
\usepackage[hidelinks]{hyperref}

\begin{document}

\title{What Does a Repair Agent Gain from Remembering Its Failure? An Exploratory Study on Real Python Bugs}
\author{\IEEEauthorblockN{Wenbo Ye}
\IEEEauthorblockA{Independent Researcher\\
ywb6113@gmail.com}
\and
\IEEEauthorblockN{Yongxi Zhou}
\IEEEauthorblockA{Independent Researcher\\
zhou.yongx@northeastern.edu}
\and
\IEEEauthorblockN{Junwei Yao}
\IEEEauthorblockA{Independent Researcher\\
yao.junw@northeastern.edu}
\and
\IEEEauthorblockN{Jiahui Ge}
\IEEEauthorblockA{Independent Researcher\\
jiahuige@alumni.usc.edu}}
\maketitle

\begin{abstract}
When a coding agent fails a repair task, it is often retried with a record of the failure: the transcript, a
self-written lesson, or extracted test signals. We ask what such failure memory changes in the retry. On 75 BugsInPy
bugs, we retried each of 181 failed attempts of \texttt{gpt-5.4-mini} under five conditions, with a predeclared
analysis whose deviations we report. Two predeclared follow-ups repeated the plain-retry, transcript and lesson conditions with the larger
\texttt{gpt-5.4} and added a
notice-only control. On the bug-revealing tests, every memory condition gained over a plain retry in point estimate
(+5.5 to +9.9 points over 29.8\%), but none significantly. Five further findings stand out. First, a model-written
lesson showed no advantage over the transcript it summarises ($-1.7$ points, 95\% CI $[-8.7, +4.9]$; $-3.5$ for
\texttt{gpt-5.4}), although it costs an extra model call. Second, the transcript may do more than tell the agent
that it failed. A notice alone produced no significant change (and was 7.1 points below a plain retry on resolution
for \texttt{gpt-5.4}), while the transcript resolved 12.2 points more failures than the notice for
\texttt{gpt-5.4-mini} (CI $[+4.2, +20.0]$, in a partial run of 156 failures; the only significant one of twelve
predeclared contrasts, and not significant for \texttt{gpt-5.4}). Third, post hoc in the main study, fewer retries
with memory ended without modifying any Python file, while the share of editing retries that touched a file the fix
changes was near ceiling in every condition. Fourth, in a post hoc check against the full test suites, retries that
had also seen the previous attempt's test results (the final output for the transcript, an extracted digest for the
signal conditions) more often failed other tests that the reference fix passes, which shrinks their apparent gains.
Fifth, outcomes varied far more between failures than between conditions, and we give the sample sizes a
confirmatory study would need. We report all exclusions, interruptions and post hoc analyses.
\end{abstract}

\section{Introduction}
Language-model agents for program repair work in a loop of inspecting files, editing code and running tests. When an
attempt ends with failing tests, a common response is to start again and pass along something from the failed
attempt. Reflexion-style agents write a verbal lesson~\cite{shinn2023reflexion}; other systems pass on the previous
trajectory, structured test output or curated experience~\cite{wang2026memgovern,fan2026vibemem}. It is not obvious
that such memory pays off. Resampling alone already recovers many failures~\cite{brown2024monkeys}, and self-repair
helps less than often assumed once its cost is counted~\cite{olausson2024selfrepair}. In ALFWorld and HumanEval,
reflexive agents have been observed to store confident but wrong readings of the task and to keep acting on them,
while programmatically extracted failure signals helped~\cite{dixit2026honest}. We know of no controlled comparison of these forms of
memory in repository-level repair, where a failed attempt leaves behind edited files and test output.

We study what failure memory changes when a repair agent retries, for a fixed harness and model family.
\textbf{(A)} Compared with a plain retry, does memory of the failure change how often the retry passes the
bug-revealing tests, and at what cost? Does summarising the failure into a lesson help compared with passing on the
transcript it was written from? What part of the effect is merely the notice that the attempt failed?
\textbf{(B)} When the model writes a lesson, where does it point, compared with the reference fix and with simple
deterministic baselines?
A plain retry is the baseline that any memory must beat, because resampling alone recovers many
failures~\cite{brown2024monkeys}. The transcript and lesson conditions receive exactly the same truncated record, so
their comparison is designed to isolate the summarisation step, and a notice-only condition separates the memory from the
announcement that the attempt failed.

The study differs from prior work in four ways. Earlier comparisons of self-diagnosis and extracted failure signals
used ALFWorld and HumanEval without a resampling baseline~\cite{dixit2026honest}; we compare transcript, lesson and
digest in repository-level repair against a plain retry, with the transcript and the lesson built from the same
record. We separate the failure notice from the memory content with a notice-only control, which we have not seen
in studies of agent retries. We check resolved retries against the full test suites, which shows that the usual
bug-revealing-test endpoint can reorder the conditions in point estimate. And we report how strongly retry outcomes cluster by
failure, which determines how such studies must be sized.

This is an exploratory evaluation, not a new method. It has three parts. A predeclared main study (M1,
\texttt{gpt-5.4-mini}, 181 failures) gave inconclusive primary contrasts. Post hoc analyses then showed how the
retries differed. Two predeclared follow-ups tested what those analyses suggested: a replication with
\texttt{gpt-5.4} (M2, 85 failures) and a notice-only control on the failures of both models (M3). Our findings are:
\begin{itemize}
\item \textbf{Lessons showed no advantage over transcripts on the bug-revealing tests.} With either model, a
model-written lesson did not outperform the transcript it summarises, despite the extra call; the upper 95\% bounds
of its advantage are +4.9 (M1) and +8.2 points (M2). On a post hoc full-suite outcome the lesson was slightly ahead.
\item \textbf{The transcript may carry more than a failure notice.} A notice alone produced no significant change.
The transcript exceeded it in point estimate for both models, but significantly only on resolution for the smaller
model, in a partial run.
\item \textbf{With memory, fewer retries stop without an edit.} Post hoc in M1, fewer retries with memory ended
without any change; in M2 this held for the transcript but not clearly for the lesson. Where editing retries edited
was near ceiling in every condition.
\item \textbf{Some apparent gains do not survive the full test suites.} Post hoc, more of the resolved retries that
had also seen the previous attempt's test output failed other tests that the reference fix passes (an association,
not a demonstrated cause).
\item \textbf{The failure matters more than the memory.} Of the 181 failures, 63 were resolved under no condition
and 17 under all five, and retry outcomes cluster by failure (ICC 0.40). Detecting effects below about 10 points
therefore needs larger task sets than ours; we give a design calculation.
\end{itemize}
For practitioners, our data show no advantage of a lesson over the transcript on the bug-revealing tests, although
advantages up to the upper bounds above are not excluded. For evaluation, comparisons of retry or memory strategies
that score only the bug-revealing tests can rank the strategies differently from a full-suite check, because
strategies that expose more test output may produce more patches that break other tests. All raw transcripts, code and analyses are released.

\section{Related Work}
\textbf{Self-repair and reflection.} Self-Debugging teaches models to repair their own code from execution
feedback~\cite{chen2023selfdebug}, and Self-Refine alternates between critiquing and revising its own
output~\cite{madaan2023selfrefine}. Olausson et al.\ find that self-repair helps less than often assumed once its cost
is counted~\cite{olausson2024selfrepair}. Repeated sampling alone raises coverage on SWE-bench
Lite~\cite{brown2024monkeys}, which is why we include a plain retry as the control. Reflexion stores verbal
reflections between trials~\cite{shinn2023reflexion}, ExpeL distils insights from experience~\cite{zhao2024expel},
and intrinsic self-correction without external feedback can fail to help~\cite{huang2024cannot}. Dixit et
al.~\cite{dixit2026honest} report memory confabulation in ALFWorld and HumanEval. In stalled ALFWorld environments,
replacing self-diagnosis with programmatically extracted failure signals raised mentions of the correct target
object from 0\% to 86\%. Our
SIGNAL condition is a repository-repair counterpart of such extraction; we did not measure confabulation as they
define it.

\textbf{Program-repair agents.} Automated program repair has a long research history~\cite{legoues2019apr}. ChatRepair carries failed patch attempts forward in a
conversation~\cite{xia2024chatrepair}. SWE-bench~\cite{jimenez2024swebench} established repository-level evaluation,
followed by agents such as SWE-agent~\cite{yang2024sweagent} and AutoCodeRover~\cite{zhang2024autocoderover} and the
pipeline approach Agentless~\cite{xia2024agentless}; RepairAgent lets an LLM agent decide autonomously which repair
tools to invoke~\cite{bouzenia2024repairagent}. Lin et al.~\cite{lin2026torun} found that restricting test
execution changed resolve rates by only 1.25 points (not significant) for commercial agents.

\textbf{Memory for coding agents.} Repository context files did not generally improve success and raised cost by over
20\%~\cite{gloaguen2026agentsmd}. Existing memory systems rarely beat a memory-off baseline, while injecting verified
experience helped four of five solvers by 1.1--4.5 points~\cite{fan2026vibemem}, and MemGovern reports gains from
governed experience cards~\cite{wang2026memgovern}. These studies concern memory across tasks; we study memory of a
failed attempt on the same task. Both the truncated transcript and the lesson are compactions of the failure record,
and compaction that serves the question at hand need not serve later ones~\cite{yao2026compaction}; here the later
question is the retry itself. We use BugsInPy~\cite{widyasari2020bugsinpy} because its bug-revealing tests can be
reproduced.

\textbf{Evaluating stochastic systems.} Randomised algorithms in software engineering should be compared over repeated
runs with suitable statistical tests~\cite{arcuri2011practical}, and code-generation benchmarks have long reported
repeated-sample metrics such as pass@$k$~\cite{chen2021codex}. LLM behaviour can change with superficial prompt
formatting~\cite{sclar2024prompt}, and LLM judges show position, verbosity and self-enhancement
biases~\cite{zheng2023judging}. We therefore keep the base prompt identical across conditions and score outcomes with
tests rather than with a judge.

\section{Study Design}
\subsection{Tasks}
We used BugsInPy bugs whose buggy and fixed versions we could build reproducibly. Their bug-revealing tests had to
fail on the buggy version and pass on the fixed version in our sandbox (checked twice each from recorded test events).
The fix commit also had to link a public issue, which we use verbatim as the task statement (no pull-request text or
diff). The issue text could not contain any line added by the reference fix, and the initial prompt had to fit
within 32{,}000 bytes.
From 501 BugsInPy bugs: native-extension projects 209, environment failures 49, sandbox control failures 16, no linked issue 129, solution leakage 6, statement too long 1, development or previously exposed 10, initial prompt above 32{,}000 bytes 6; 75 tasks remain.

The ten development or previously used tasks are the eight pilot tasks of this study and two tasks inspected while developing
the harness. The selection
rules were fixed before screening and do not depend on model outcomes. The 75 tasks come from 12 projects (ansible,
black, fastapi, httpie, keras, luigi, sanic, scrapy, thefuck, tornado, tqdm, youtube-dl); the selected tests are the
bug-revealing tests listed by BugsInPy (revision \texttt{11c5f1e}).

\subsection{Agent and harness}
The agent is \texttt{gpt-5.4-mini-2026-03-17} (temperature 1, low reasoning effort, at most 3{,}000 output tokens per
response, no sampling seed). Each response must be a single JSON operation: run a shell command, run the selected
tests, or declare the task done. An attempt may use at most eight responses. Commands run in a sandbox without
network access, with test files read-only and a 60-second limit per command, and the agent always sees the full test
output. The initial prompt contains the issue text and the output of one test run on the buggy code. An attempt also
ends early in two cases: when the next request would exceed a conservative estimate of 96{,}000 input bytes (``input
bound reached''), and when a response uses its whole output budget without producing an operation (``output budget
exhausted''). After an
attempt ends, the selected tests are run once more and scored from recorded pytest events; the attempt counts as
resolved only if all of them pass. Attempts that add pytest configuration, a \texttt{conftest.py} or modules that
shadow the test runner are flagged: flagged first attempts are excluded and flagged retries count as unresolved (none
occurred). All raw model requests, responses, tool logs and workspace snapshots were retained. Prompts, schemas and code are in
the replication package.

\subsection{Retry conditions}
Each task received six independent first attempts. Every first attempt that did not pass the final evaluation, for
whatever agent-caused reason (including broken builds, timeouts and limits), is a \emph{failure}. Each failure
received five retries from the original workspace, one per condition. They ran one after another, in an order rotated
by task and repetition, and tasks ran in parallel on 4--6 workers:
\begin{itemize}
\item \textbf{NONE}: the first-attempt prompt, unchanged.
\item \textbf{HISTORY}: the failure transcript appended, with each message truncated to 2{,}000 characters (head and
tail) and the final test output to 6{,}000 characters.
\item \textbf{REFLECT}: a lesson written by the same model with the same sampling settings. The model sees the task
statement and the same truncated failure record as HISTORY, and fills a fixed schema (suspected files, suspected
functions, root cause, what failed, next steps).
\item \textbf{SIGNAL}: a deterministic digest listing the final status, the failing tests, up to 30 assertion or
exception lines (3{,}000 characters) and the files the failed attempt modified.
\item \textbf{BOTH}: REFLECT and SIGNAL together.
\end{itemize}
No condition sees the reference fix. HISTORY passes on the truncated record and REFLECT passes on the model's
summary of the same record, so their difference reflects the summarisation step. SIGNAL is a separate, deterministic
digest. Fig.~\ref{fig:example} shows the memory each condition received for one failure.

\begin{figure*}[t]
\centering\footnotesize
\fbox{\begin{minipage}{0.97\textwidth}\raggedright
\textbf{Shared header (all memory conditions).} \texttt{INFORMATION FROM A PREVIOUS ATTEMPT ON THIS TASK. That attempt did
not make the selected tests pass. The workspace has been reset to its original state.}\\[3pt]
\textbf{HISTORY.} \texttt{Record of the previous attempt: FAILED ATTEMPT: --- AGENT OPERATION 1 ---
\{"action":"done"\} FINAL EVALUATION OUTPUT OF THE SELECTED TESTS: [pytest log including] E RuntimeError: current
IOLoop already exists \,|\, tornado/ioloop.py:253: RuntimeError}\\[3pt]
\textbf{REFLECT.} \texttt{Notes written after the previous attempt: Suspected files: tornado/ioloop.py,
tornado/test/ioloop\_test.py. Suspected functions: IOLoop.initialize, TestIOLoopCurrent.test\_force\_current. Root
cause: The failure output shows an inverted existence check in IOLoop.initialize(make\_current=True) [\ldots] I cannot
confirm the exact surrounding logic without the source [\ldots] What failed: [\ldots] Next steps: Inspect
tornado/ioloop.py around IOLoop.initialize and fix the make\_current=True branch [\ldots]}\\[3pt]
\textbf{SIGNAL.} \texttt{Observed results of the previous attempt: Final evaluation status: test\_failures\_observed.
Failing selected tests: tornado/test/ioloop\_test.py::TestIOLoopCurrent::test\_force\_current (call). Error lines from
the final test output: E RuntimeError: current IOLoop already exists. Files the previous attempt modified: none.}
\end{minipage}}
\caption{Memory appended to the retry prompt for one failure (tornado/14, repetition 6; abridged, [\ldots] marks
omissions). The first attempt declared the task done without changing any file. BOTH appends the REFLECT and SIGNAL
blocks; NONE appends nothing. In this instance all five retries resolved the bug.}
\label{fig:example}
\end{figure*}

\subsection{Outcomes and analysis}
The unit of analysis is a failure (task, repetition), clustered by task, since repeated runs on the same item are
correlated and pooled-binomial intervals over them misstate uncertainty~\cite{zhou2026repeated}. The primary outcome is
whether the retry passes, by intention to treat: retries that end unresolved for any agent-caused reason count as
failures, and retries lost to harness errors are treated as missing. We compare each memory condition with NONE using
mean paired differences, 95\% task-cluster bootstrap intervals~\cite{cameron2008bootstrap} (10{,}000 resamples, seed
20261015), task-level sign-flip permutation tests and Holm correction~\cite{holm1979simple}. We interpret the results
mainly through the intervals. With only
11 projects contributing failures we did not cluster by project and instead report a leave-one-project-out analysis.
Because attempts are stochastic, NONE measures what simple resampling recovers, the baseline any memory must
beat. The planned secondary comparisons are REFLECT vs HISTORY, REFLECT vs SIGNAL and BOTH vs REFLECT, plus cost and
wall-clock time. The planned sensitivity analyses are a complete-case analysis, task-weighted estimates, an analysis
excluding late-started tasks, imputation of missing retries, and results by execution batch (batches 1--3 are the
tasks retained from the three runs that produced data; Section~\ref{sec:threats}). Analyses marked \emph{post hoc} were added after the data were
seen. Further breakdowns by batch are in the replication package.

For (B), we match the locations named in each lesson against the source files and functions changed by the reference
fix. Test files are not part of the reference, so naming a test file (visible in the failure output) never counts as a
hit and lowers precision. Matching uses a lenient rule (path suffix, unqualified name), a strict rule (exact path, qualified name) and a first-listed
rule. We compare them with two deterministic baselines: the locations the failed attempt itself edited, and the
source frames in its final traceback (mapped to the enclosing function). We also flag lessons that mention a test or
exception that does not appear in the failure record. This check does not say whether a diagnosis is correct. Both
(A) and (B) are scored by fixed rules rather than by an LLM judge, because judge verdicts in safety evaluation change
with content-invariant framing of the judged text~\cite{zhou2026style}.

\subsection{Replication with a stronger model}\label{sec:m2design}
After the post hoc analyses of Section~\ref{sec:mechanism}, we ran a smaller replication (M2) with
\texttt{gpt-5.4-2026-03-05}, the larger model of the same family. All other settings were unchanged: the same 75
tasks and inputs and six first attempts per task, with retries under NONE, HISTORY and REFLECT only. Its protocol was
frozen and reviewed before any M2 call. The primary hypothesis (H1) came from the M1 post hoc results: memory raises
the share of retries that modify at least one Python file, for both HISTORY and REFLECT. H1 counts as supported only
if both contrasts are positive with Holm-adjusted $p<0.05$; otherwise it is inconclusive. Retry
resolution and REFLECT $-$ HISTORY are secondary and reported as estimates. M2 is never pooled with M1, and it
reuses the tasks that generated its hypothesis, so it tests transfer to a stronger model, not to new tasks.

\subsection{Notice-only control}\label{sec:m3design}
Every memory condition begins with a notice that the previous attempt failed and the workspace was reset, which NONE
lacks (Fig.~\ref{fig:example}). To separate the notice from the memory, we ran a further predeclared follow-up (M3). Every M1 and M2 failure
received four new retries with that failure's model, run in the same session from the original workspace: NONE,
NOTICE (the task prompt plus exactly the notice, without any memory), and HISTORY and REFLECT with the same prompts,
byte for byte, as in M1/M2. The contrasts therefore compare runs made together, not weeks apart. The predeclared hypothesis N1 is that NOTICE
$-$ NONE is positive on modifying a Python file, that is, that the notice alone reduces retries without an edit. For each model, the
primary family is NOTICE $-$ NONE, HISTORY $-$ NOTICE and REFLECT $-$ NOTICE on modifying at least one Python file
(Holm over three). The same contrasts on resolution form a secondary family. Before any run we fixed a decision
table. A contrast is \emph{positive} or \emph{negative} only if significant after Holm correction. A content contrast
(HISTORY or REFLECT $-$ NOTICE) is \emph{equivalent} if its interval lies within $\pm$10 points. Anything else is
\emph{not distinguished}. The protocol, code and both resumptions described in Section~\ref{sec:threats} were
reviewed before the corresponding runs.

\section{Results}
\subsection{First attempts}
Of 450 first attempts, one was lost to a provider error (HTTP~400, cause unknown). Of the other 449, 268 passed and
181 failed. The failures came from 52 tasks in 11 of the 12 projects (keras had none), with one to six failures per
task. Table~\ref{tab:subtypes} shows how attempts ended: 15 first attempts exhausted the output budget and 17 reached
the input bound. Every failure produced a valid lesson. One REFLECT retry was also lost to a provider error
(HTTP~400). All truncated responses stopped at the length limit; none were refusals.

\begin{table*}[t]
\caption{How all first attempts and retries ended. ``Of which resolved'' counts attempts whose final evaluation
passed, regardless of how they ended.}
\label{tab:subtypes}
\centering\small
\begin{tabular}{lrrrrrr}
\toprule
Terminal status (all attempts) & First & NONE & HISTORY & REFLECT & SIGNAL & BOTH \\
\midrule
declared done & 280 & 70 & 76 & 60 & 70 & 61 \\
eight responses used & 131 & 89 & 79 & 106 & 89 & 104 \\
output budget exhausted & 15 & 12 & 8 & 6 & 10 & 6 \\
input bound reached & 17 & 6 & 11 & 5 & 9 & 8 \\
invalid test run & 6 & 4 & 7 & 3 & 3 & 2 \\
provider error (missing) & 1 & 0 & 0 & 1 & 0 & 0 \\
\midrule
of which resolved & 268 & 54 & 71 & 67 & 64 & 72 \\
\bottomrule
\end{tabular}

\end{table*}

\subsection{Does retry memory help? (A)}
\begin{figure}[t]
\centering
\includegraphics[width=\columnwidth]{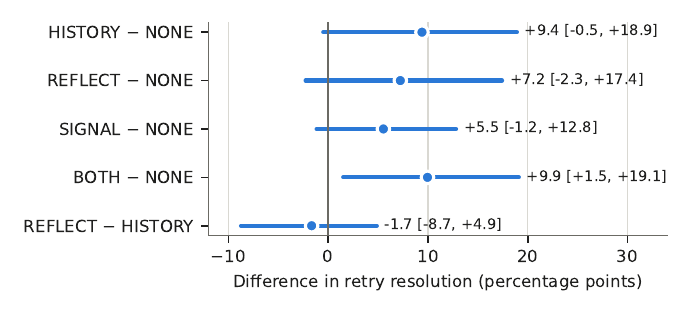}
\caption{Paired differences in retry resolution (percentage points, 95\% task-cluster bootstrap intervals).
REFLECT $-$ HISTORY compares a model-written lesson with the transcript it was written from.}
\label{fig:contrasts}
\end{figure}

Table~\ref{tab:primary} and Fig.~\ref{fig:contrasts} show the primary analysis. A plain retry fixed 29.8\% of
failures. Every memory condition had a higher point estimate, from +5.5 points (SIGNAL) to +9.9 points (BOTH). Weighting
tasks equally gives similar values (+8.1 to +11.4). None of the differences is significant after Holm
correction ($p_{\mathrm{Holm}}$ 0.17--0.33). Unadjusted, the BOTH interval excludes zero, and the complete-case HISTORY contrast below has $p=0.013$. The
intervals are not adjusted for multiplicity, so we do not read these as evidence of a benefit. They range from
$-2.3$ to +19.1 points: they exclude more than a small harm and gains above about 19 points, but cannot separate a
modest gain from none. Resampling projects instead of tasks (11
clusters, post hoc) gives similar intervals, from $-2.9$ to +24.8 points.

\begin{table*}[t]
\caption{Primary analysis: share of the 181 failures (from 52 of the 75 tasks) resolved by the retry, by condition.
Differences ($\Delta$) are paired, in percentage points, with task-cluster bootstrap intervals.}
\label{tab:primary}
\centering\small
\setlength\tabcolsep{5pt}\begin{tabular}{lrrrcrr}
\toprule
Condition & Resolved & $n$ & $\Delta$ (pp) & 95\% CI & $p_{\mathrm{Holm}}$ & Task-weighted $\Delta$ \\
\midrule
NONE & 29.8\% & 181 & -- & -- & -- & -- \\
HISTORY & 39.2\% & 181 & +9.4 & [$-$0.5, +18.9] & 0.24 & +11.4 \\
REFLECT & 37.2\% & 180 & +7.2 & [$-$2.3, +17.4] & 0.33 & +8.1 \\
SIGNAL & 35.4\% & 181 & +5.5 & [$-$1.2, +12.8] & 0.33 & +9.6 \\
BOTH & 39.8\% & 181 & +9.9 & [+1.5, +19.1] & 0.17 & +11.3 \\
\bottomrule
\end{tabular}

\end{table*}

\textbf{Lesson versus transcript.} REFLECT and HISTORY are built from the same record. Their difference is $-1.7$
points (95\% CI $[-8.7, +4.9]$), so we see no benefit from summarising. The interval allows lessons to be up to about
nine points worse or five points better. The other planned secondary comparisons are also small and imprecise: REFLECT
vs SIGNAL +1.7 $[-5.7, +9.2]$ and BOTH vs REFLECT +2.2 $[-3.9, +8.7]$. A plain retry cost about \$0.021 per failure
on average. HISTORY added about \$0.004 per retry, REFLECT \$0.009 and BOTH \$0.008 (including the lesson call). REFLECT and
BOTH also took about 21--23 seconds longer.

\textbf{Sensitivity, heterogeneity and the interim look.} Table~\ref{tab:sens} gives the sensitivity analyses; all
of them keep every point estimate positive, unlike the batch rows of the same table. Two of them are not counted as support. Dropping retries that reached the input bound (HISTORY +13.3, unadjusted
$p=0.013$) favours the memory conditions by construction, because this happened more often with HISTORY (11 vs 6
retries). Counting those retries as failures, which reclassifies four that had passed, also raised the estimates
(+6.1 to +10.5). Excluding black, the largest
project (62 of 181 failures), gives +5.9 to +10.1, and leaving out any single project gives +3.5 to +13.0.
Excluding late-started tasks gives +9.7, +7.4, +5.1 and +10.2, and imputing the missing REFLECT retry as passed
or failed gives +7.7 or +7.2.
The estimates differ sharply between the three execution batches, which are alphabetical groups of projects with
plain-retry rates of 17\%, 28\% and 46\%; we do not interpret this. Because of the run interruptions
(Section~\ref{sec:threats}), the first 46 tasks finished first. An interim analysis of them gave HISTORY +18.6
($[+9.0, +29.9]$, $p_{\mathrm{Holm}}=0.007$; Table~\ref{tab:sens}), and we then decided to complete the planned 75 tasks, as recorded beforehand but after
seeing the result. The remaining 29 tasks showed no gain ($-5.9$ to $+4.4$), and over all 75 tasks HISTORY is +9.4
($p_{\mathrm{Holm}}=0.24$). The target, primary outcome and contrast family were unchanged, no alpha-spending
adjustment was made, and all $p$-values are nominal. Per-project results are in the replication package.

\begin{table*}[t]
\caption{Differences versus NONE (percentage points) under alternative analyses.
$^a$Analysed before the decision to complete the task set (interim look; nominal $p$). $^b$Drops retries that reached
the input bound or were flagged; biased toward the memory conditions. $^c$Post hoc. Batches are alphabetical groups
of projects.}
\label{tab:sens}
\centering\small
{\setlength\tabcolsep{8pt}\begin{tabular}{lrrrr}
\toprule
Analysis & HISTORY & REFLECT & SIGNAL & BOTH \\
\midrule
Primary (75 tasks) & +9.4 & +7.2 & +5.5 & +9.9 \\
First look (46 tasks)$^{a}$ & +18.6 & +11.5 & +8.8 & +13.3 \\
Complete case$^{b}$ & +13.3 & +7.6 & +7.7 & +11.9 \\
Excluding black$^{c}$ & +8.4 & +8.5 & +5.9 & +10.1 \\
Leave one project out$^{c}$ & +5.8--13.0 & +3.5--8.7 & +3.5--7.5 & +6.4--11.6 \\
Batch 1 (24 tasks) & +13.6 & +7.4 & +7.4 & +8.6 \\
Batch 2 (22 tasks) & +31.2 & +21.9 & +12.5 & +25.0 \\
Batch 3 (29 tasks) & $-$5.9 & 0.0 & 0.0 & +4.4 \\
\bottomrule
\end{tabular}
}
\end{table*}

\subsection{How does memory change the retry? (post hoc)}\label{sec:mechanism}
We added the analyses in this subsection after seeing the data, to examine how memory changes the retry. Their
intervals are not adjusted for multiplicity.

\textbf{With memory, fewer retries end without an edit.} For each retry we recorded whether it edited any source file (any Python file; test files are read-only) and
whether it edited a file changed by the reference fix (Table~\ref{tab:mechanism}). A plain retry edited no source
file in 30.9\% of failures (56). Of these, 22 declared the task done without changing anything, 21 ran out of
responses, 11 exhausted the output budget and 2 reached the input bound. With the transcript the share fell to
13.8\% (25), mainly because only 4 retries declared the task done without an edit. Every memory condition begins with
a header stating that the previous attempt did not pass (Fig.~\ref{fig:example}). NONE has no such header: a plain
retry receives exactly the first-attempt prompt. The header alone could therefore explain the drop, and the
notice-only control in Section~\ref{sec:m3} tests this. Every memory condition raised the share of retries that
edited a source file, most clearly HISTORY (+17.1 points, 95\% CI $[+7.9, +26.1]$) and REFLECT (+13.3,
$[+6.3, +20.6]$).

Memory did not improve \emph{where} the retry edited. In every condition, 94--98\% of the retries that edited
anything edited a fix file, which is unsurprising: the reference fix changes a single source file in 45 of the 52
tasks with failures. Retries that left the fix files untouched practically never passed (1 of 215 across
conditions). Among retries that did edit a fix file, the pass rate was 44.3\% for NONE and 46.4--53.0\% with memory.
These differences from NONE are small and imprecise (HISTORY +2.1, $[-9.5, +13.5]$; BOTH +8.7, $[-2.4, +19.9]$;
unpaired subsets, task-cluster bootstrap). The gain in point estimates in Table~\ref{tab:primary} is about what the
higher rate of editing would predict (for HISTORY, $0.845 \times 0.464 = 0.39$ versus $0.674 \times 0.443 = 0.30$). Editing is itself an outcome
of the retry, so this decomposition is descriptive, not a causal mediation analysis.

\begin{table*}[t]
\caption{How the retry changed (post hoc): share of retries that edited any source file and a file changed by the
reference fix, paired difference versus NONE (task-cluster bootstrap), and share resolved among retries that did or
did not edit a fix file. The last two columns condition on a post-treatment variable and are descriptive.}
\label{tab:mechanism}
\centering\small
\setlength\tabcolsep{4pt}\begin{tabular}{lrcrrr}
\toprule
Condition & Edited source & $\Delta$ vs NONE (pp) & Fix file among & \multicolumn{2}{c}{Resolved (\%) if fix file} \\
 & (\%) & 95\% CI & editing (\%) & edited ($n$) & not edited ($n$) \\
\midrule
NONE & 69.1 & -- & 97.6 & 44.3 (122) & 0.0 (59) \\
HISTORY & 86.2 & +17.1 [+7.9, +26.1] & 98.1 & 46.4 (153) & 0.0 (28) \\
REFLECT & 82.2 & +13.3 [+6.3, +20.6] & 95.9 & 47.2 (142) & 0.0 (38) \\
SIGNAL & 79.0 & +9.9 [+2.0, +18.0] & 96.5 & 46.4 (138) & 0.0 (43) \\
BOTH & 78.5 & +9.4 [+1.2, +17.7] & 94.4 & 53.0 (134) & 2.1 (47) \\
\bottomrule
\end{tabular}

\end{table*}

\textbf{The failure matters more than the condition.} Each failure was retried under all five conditions. In 63 of the
181 failures no condition succeeded, and in 17 all five did; the remaining 101 were split. The intraclass correlation
of retry outcomes within a failure is 0.40 (0.37 within a task), whereas the condition accounts for 0.6\% of the
outcome variance; this share is necessarily small for a binary outcome whose condition means lie between 30\% and
40\%, so we rely mainly on the intraclass correlation. Most of the variation in retry success therefore lies between failures. This is why the contrasts are imprecise
even with paired retries, and why the batch and project differences in Table~\ref{tab:sens} are large.

\textbf{Do resolved retries break other tests?} Our primary outcome uses only the bug-revealing tests. Post hoc, we
ran every resolved retry and every resolved first attempt (of tasks with failures) against the full test suite of its
project, in the same sandbox. We compared each run with two runs of the fixed version. Because the baseline is the fixed version, not the buggy
one, a failing test need not be a regression the patch introduced; for brevity we say that a patch \emph{broke other
tests} if a test that passed in both reference runs did not pass with the patch. Test files were taken from the fixed
version, and the rule was written before any patch was run. Two implementation amendments, also recorded before any
patch was run, handle pytest file naming and a shutdown hang in luigi. One luigi task has no usable reference, so 11
resolved attempts are unchecked. Table~\ref{tab:fullsuite} shows the result. Few plain or lesson-based retries broke
other tests (3 of 54 and 3 of 67). Retries that saw the transcript or the digest of failing tests did so more often
(HISTORY 11 of 71, SIGNAL 9 of 64, BOTH 16 of 72), spread over 7--10 tasks per condition. Counting only retries that
pass the full suite, the estimates fall to +5.0 (HISTORY), +2.2 (SIGNAL) and +2.8 (BOTH) points, while REFLECT stays at
+7.2; all intervals include zero. On this outcome the lesson is slightly ahead of the transcript in point estimate,
a gap that almost vanishes in the sensitivity analysis below (+6.7 vs +6.1). Many of the broken
tests in black check that \texttt{black.py} is formatted by black itself, and some tqdm tests measure timing. As a
sensitivity analysis chosen after seeing the results, we also report an outcome that ignores these two kinds of
test; the pattern weakens but remains (Table~\ref{tab:fullsuite}). Every retry, including NONE, sees one run of the
selected tests on the buggy code in its prompt. HISTORY adds the failed attempt's final test output and its edits, SIGNAL and BOTH add an
extracted digest of that output, and REFLECT passes only the model's summary. One possible explanation is that seeing how a
previous patch failed the visible tests invites patches aimed at those tests, as observed for program repair
generally~\cite{smith2015cure}. The counts are small and the analysis is post hoc, so this remains a hypothesis.

\begin{table*}[t]
\caption{Full-suite check of resolved retries (post hoc). ``Broke other tests'': a test that passed in both runs of the
fixed version failed with the patch. Suite-clean: resolved on the selected tests and broke no other test (11
attempts of one luigi task without a usable reference keep their selected-test outcome). Differences are paired,
with task-cluster bootstrap intervals. $^f$Ignoring black's self-formatting checks and tqdm's timing tests
(sensitivity chosen after seeing the results).}
\label{tab:fullsuite}
\centering\small
\setlength\tabcolsep{4pt}\begin{tabular}{lrrrcc}
\toprule
Condition & Resolved & Broke other & Suite-clean & $\Delta$ vs NONE (pp) & $\Delta$ vs NONE (pp) \\
 & retries & tests & (\%) & all tests & behavioural tests$^f$ \\
\midrule
NONE & 54 & 3 & 28.2 & -- & -- \\
HISTORY & 71 & 11 & 33.1 & +5.0 [$-$4.3, +14.3] & +6.1 [$-$3.4, +15.7] \\
REFLECT & 67 & 3 & 35.6 & +7.2 [$-$2.1, +17.0] & +6.7 [$-$2.7, +16.6] \\
SIGNAL & 64 & 9 & 30.4 & +2.2 [$-$3.8, +8.8] & +3.9 [$-$2.5, +10.7] \\
BOTH & 72 & 16 & 30.9 & +2.8 [$-$5.6, +10.9] & +4.4 [$-$3.6, +12.7] \\
\bottomrule
\end{tabular}

\end{table*}

\subsection{Does the pattern hold for a stronger model?}\label{sec:m2}
Table~\ref{tab:m2} compares M2 with M1. The larger model resolved 365 of 450 first attempts, leaving 85 failures in
32 tasks, under half as many as M1. All 75 tasks completed without harness errors, every failure produced a lesson,
and M2 cost about US\$54 including a two-task smoke test. With HISTORY, more retries modified a Python file than with
a plain retry (+14.1 points, 95\% CI $[+5.1, +24.1]$, $p_{\mathrm{Holm}}=0.033$), close to the M1 estimate. With
REFLECT the difference was smaller and not significant (+5.9, $[-2.4, +16.4]$, $p_{\mathrm{Holm}}=0.37$), so the
predeclared H1 is \emph{not supported}. For the transcript the M1 pattern recurred. For the lesson the result is
inconclusive, since its interval also contains the M1 estimate (+13.3). The notice-only control
(Section~\ref{sec:m3}) also covers these failures. As in M1, 94--95\% of editing retries edited a fix file in every
condition. Among retries that edited a fix file, the pass rates were 56.1\% (NONE), 59.7\% (HISTORY) and 60.6\%
(REFLECT; HISTORY $-$ NONE +3.7, $[-19.7, +26.5]$). Retry resolution again favoured memory in point estimate (HISTORY
+10.6, REFLECT +7.1), with intervals of about $\pm$19 points. Again we found no evidence that the lesson beats the
transcript ($-3.5$, $[-14.1, +8.2]$). A plain retry cost US\$0.084
per failure, HISTORY US\$0.095 and REFLECT US\$0.115 including the lesson call. We did not run the full-suite check
on M2.

\begin{table}[t]
\caption{M1 and the M2 replication side by side (differences in percentage points with 95\% task-cluster bootstrap
intervals; not pooled, not tested against each other). $^g$Post hoc in M1; predeclared primary hypothesis in M2. $^h$NONE / HISTORY / REFLECT; in M1, SIGNAL and BOTH
give 96.5 / 94.4 and 46.4 / 53.0 (Table~\ref{tab:mechanism}).}
\label{tab:m2}
\centering\scriptsize
\setlength\tabcolsep{4pt}\begin{tabular}{lcc}
\toprule
 & M1: \texttt{gpt-5.4-mini} & M2: \texttt{gpt-5.4} \\
\midrule
First attempts resolved & 268 / 449 & 365 / 450 \\
Failures (tasks with failures) & 181 (52) & 85 (32) \\
Plain retry resolved (\%) & 29.8 & 43.5 \\
Edited a .py file: HISTORY $-$ NONE$^g$ & +17.1 [+7.9, +26.1] & +14.1 [+5.1, +24.1] \\
Edited a .py file: REFLECT $-$ NONE$^g$ & +13.3 [+6.3, +20.6] & +5.9 [$-$2.4, +16.4] \\
Fix file among editing retries (\%)$^h$ & 97.6 / 98.1 / 95.9 & 94.3 / 93.9 / 94.7 \\
Resolved if fix file edited (\%)$^h$ & 44.3 / 46.4 / 47.2 & 56.1 / 59.7 / 60.6 \\
Resolved: HISTORY $-$ NONE & +9.4 [$-$0.5, +18.9] & +10.6 [$-$8.9, +29.3] \\
Resolved: REFLECT $-$ NONE & +7.2 [$-$2.3, +17.4] & +7.1 [$-$9.2, +22.7] \\
Resolved: REFLECT $-$ HISTORY & $-$1.7 [$-$8.7, +4.9] & $-$3.5 [$-$14.1, +8.2] \\
\bottomrule
\end{tabular}

\end{table}

\subsection{Is it the notice or the memory?}\label{sec:m3}
Table~\ref{tab:m3} gives the notice-only control. For neither model was the effect of the notice alone significant. On modifying a Python file, NOTICE $-$ NONE was
+1.9 points ($[-6.3, +9.9]$) for \texttt{gpt-5.4-mini} and +5.9 ($[-7.8, +19.5]$) for \texttt{gpt-5.4}; on resolution
it was close to zero or negative. The predeclared N1 is therefore not supported for either model. The transcript exceeded the notice in every point estimate. HISTORY $-$ NOTICE was +12.8 ($[+3.3, +22.4]$) and
+15.3 ($[+3.0, +28.4]$) points on modifying a file, not significant after Holm correction ($p_{\mathrm{Holm}}$ 0.07
and 0.36). On resolution it was +12.2 ($[+4.2, +20.0]$, $p_{\mathrm{Holm}}=0.033$) and +16.5 ($[+0.0, +33.0]$)
points. The first of these is the only one of the twelve contrasts in Table~\ref{tab:m3} classed as positive. Some
bootstrap intervals exclude zero while the Holm-adjusted permutation $p$-values do not reach 0.05; with 32--46 task
clusters the two methods can disagree, and our decisions follow the predeclared permutation tests. REFLECT
fell in between and was not distinguished from NOTICE (+5.9 to +7.7 points). The predeclared equivalence label applied to the content contrasts (HISTORY or REFLECT $-$ NOTICE), and none of
them lies within $\pm$10 points. The smaller model's NOTICE $-$ NONE interval on editing does lie within $\pm$10
points, so a large notice effect is unlikely there; for \texttt{gpt-5.4} it cannot be excluded. The data therefore
do not support the reading that the memory conditions work merely by announcing the failure, at least for the
transcript, but they do not establish the size of the transcript's own contribution either. In M3, \texttt{gpt-5.4} never
declared a task done without an edit in any condition. The plain-retry rates (28\% and 48\%) differ from M1 and M2,
as expected for new samples. The mini part stopped near its end on an infrastructure error, so 156 of 181 failures
were analysed; the \texttt{gpt-5.4} part was resumed twice after interruptions (Section~\ref{sec:threats}).
M3 cost US\$48.8.

\begin{table*}[t]
\caption{Notice-only control (M3): paired differences in percentage points with 95\% task-cluster bootstrap intervals
and the predeclared decision (n.s.: N1 not supported; n.d.: not distinguished; positive: $p_{\mathrm{Holm}}<0.05$).
Holm over the three contrasts within each outcome and model. The predeclared equivalence label ($\pm$10 points)
applies to the HISTORY and REFLECT $-$ NOTICE contrasts only. $^i$Partial: 46 of 52 tasks.}
\label{tab:m3}
\centering\small
\setlength\tabcolsep{3pt}\begin{tabular}{lcc}
\toprule
 & \texttt{gpt-5.4-mini} (156 failures)$^i$ & \texttt{gpt-5.4} (85 failures) \\
\midrule
Modified a .py file: NOTICE $-$ NONE & +1.9 [$-$6.3, +9.9] (n.s.) & +5.9 [$-$7.8, +19.5] (n.s.) \\
Modified a .py file: HISTORY $-$ NOTICE & +12.8 [+3.3, +22.4] (n.d.) & +15.3 [+3.0, +28.4] (n.d.) \\
Modified a .py file: REFLECT $-$ NOTICE & +7.7 [$-$1.3, +16.9] (n.d.) & +5.9 [$-$4.6, +19.0] (n.d.) \\
Resolved: NOTICE $-$ NONE & +3.2 [$-$5.0, +11.9] (n.s.) & $-$7.1 [$-$19.7, +5.2] (n.s.) \\
Resolved: HISTORY $-$ NOTICE & +12.2 [+4.2, +20.0] (positive) & +16.5 [+0.0, +33.0] (n.d.) \\
Resolved: REFLECT $-$ NOTICE & +6.4 [$-$1.4, +14.4] (n.d.) & +5.9 [$-$4.7, +15.6] (n.d.) \\
\midrule
Resolved (\%): NONE / NOTICE / HISTORY / REFLECT & 28 / 31 / 43 / 37 & 48 / 41 / 58 / 47 \\
\bottomrule
\end{tabular}

\end{table*}

\subsection{Where do the lessons point? (B)}
\begin{figure}[t]
\centering
\includegraphics[width=\columnwidth]{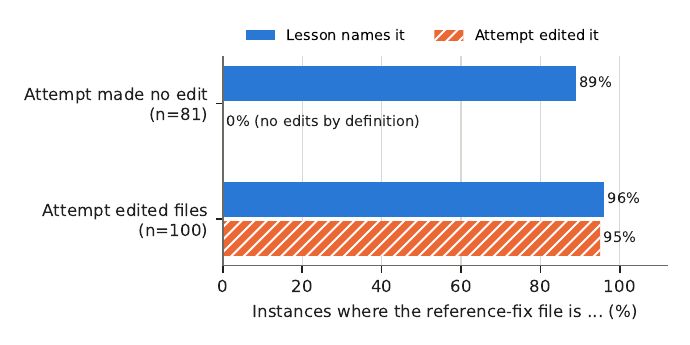}
\caption{Share of failures in which the lesson names a file changed by the reference fix, and in which the failed
attempt had edited one, split by whether the attempt edited any source file (post hoc).}
\label{fig:lessons}
\end{figure}

Lessons usually name the right file. When the failed attempt had edited files, they rarely improved on it; their
gain comes from failures in which the attempt edited nothing.
They named a file changed by the reference fix in 92.8\% of failures (first-listed 87.3\%; Table~\ref{tab:lessons})
and a changed function in 75.1\% of the 169 failures whose fix changes a function. The deterministic baselines score
lower overall (attempt edits: 52.5\% file; traceback frames: 22.7\%), but 81 of the 181 failed attempts edited no
source file. Fig.~\ref{fig:lessons} (post hoc) separates the two cases. Where the attempt edited files ($n=100$),
lessons named a changed file in 96\% of cases and the attempt's own edits in 95\%, while lessons listed more
candidates (2.3 vs 1.0 files) with lower precision (0.46 vs 0.94). Almost all of the lessons' additional correct
files (72 of 73) come from failures in which the attempt made no edit. At function level the lessons named a changed
function more often (83.5\% vs 74.7\%, $n=91$), but with longer candidate lists and no advantage for the first-listed
function (61.5\% vs 62.6\%). File-level hits are near ceiling outside black (99.2\%), so function-level hits and
precision are the more informative measures.

Post hoc, a lesson that pointed to the right file was at best weakly associated with success: REFLECT retries
passed in 37.6\% of failures whose lesson listed a changed file first and 34.8\% otherwise ($n=157$ and 23), and
in those failures REFLECT did not beat HISTORY ($-1.9$, $[-8.2, +4.3]$). Seven lessons (3.9\%) mentioned a test or
exception absent from the failure record. We did not assess whether the diagnoses themselves are correct; a blind
annotation protocol and 60 sampled lessons are in the replication package.

\begin{table*}[t]
\caption{Where the reference fix is, according to the lesson and to deterministic baselines (181 failures).
$^d$On the 169 failures whose fix changes a function.}
\label{tab:lessons}
\centering\small
\begin{tabular}{lrr}
\toprule
Location source & File hit & Function hit$^{d}$ \\
\midrule
Lesson (any listed) & 92.8\% & 75.1\% \\
Lesson (first listed) & 87.3\% & 55.6\% \\
Attempt's own edits & 52.5\% & 40.2\% \\
Traceback frames & 22.7\% & 13.0\% \\
\midrule
\multicolumn{3}{l}{Post hoc, attempt edited files (n=100): lesson 96\% vs attempt 95\% (file)} \\
\multicolumn{3}{l}{Same subset: listed files 2.3 vs 1.0; file precision 0.46 vs 0.94} \\
\bottomrule
\end{tabular}

\end{table*}

\section{Discussion}
These data cannot tell us whether failure memory raises the pass rate of a retry. The intervals run from a small harm
to gains of about 19 points over a 29.8\% baseline. Point estimates are positive for every condition and in every
leave-one-project-out analysis, though not in every batch, and the largest come from the second of the three
batches.

The post hoc analyses describe how the retries differ. With memory, fewer retries ended without modifying any Python
file, chiefly because fewer declared the task done unchanged; the others without an edit ran out of responses or
output budget. Each memory condition is a package: it adds a notice that the previous attempt failed as well as the
memory itself. The notice-only control speaks against the notice being the whole story for the transcript. The notice
alone had no significant effect for either model, and the transcript exceeded it in every point estimate,
significantly only for the smaller model's resolution. For the larger model a notice effect cannot be excluded, and
for the lesson the comparison with the notice was not distinguished. The plausible difference is the concrete record
of what was tried and how it failed, which the transcript carries and the notice does not.

Memory did not make retries edit in better places, since almost every editing retry already touched a fix file.
Among retries that did, the pass rates with memory were only slightly and imprecisely higher. The \texttt{gpt-5.4}
replication showed the same pattern for the transcript and was inconclusive for the lesson. The larger model also
ended fewer plain retries without an edit (17.6\% vs 30.9\%), which leaves less room for memory to help. Part of this
pattern may be specific to our tight harness, where many attempts end by running out of responses; with a larger
budget, a plain retry may give up less often. The lesson was written by the same small model at low reasoning
effort, and a stronger lesson writer might do better; we did not test either.

For practitioners using a model and harness like ours, a lesson did not raise the pass rate on the selected tests
over passing on the transcript (upper 95\% bounds +4.9 points in M1 and +8.2 in M2), and it costs an extra call. On the
full-suite outcome the lesson was slightly ahead, so the choice between the two is open; what our data do support is
checking retries that saw failing tests against a wider suite. For localisation, when the failed attempt had edited
files, lessons named a changed file about as often as those edits did while listing more candidates (higher recall
at function level, much lower precision). Our results do not bear on confabulation findings in other
environments~\cite{dixit2026honest}.

\textbf{How large a study would be needed?} We ran a post hoc design calculation based on the task-cluster
standard errors of the primary contrasts (0.036--0.050 with 52 tasks contributing failures). Detecting a 7-point
gain with 80\% power at $\alpha=0.05$ would require about 106--210 tasks with failures, and 150--299 at the
$\alpha=0.0125$ used by the first Holm step; for a 10-point gain the figures are 52--103 and 74--146. These figures are approximate, because they assume that
the observed standard errors carry over to a larger task sample. About 70\% of our tasks produced failures, so a
confirmatory study of a 7-point effect would need roughly two to six times the present task set. Because outcomes
cluster by failure, adding retries per failure helps less than adding tasks~\cite{zhou2026repeated}. Future studies should also sample
tasks from more projects, since project blocks differed in their plain-retry rates.

\section{Threats to Validity}\label{sec:threats}
\textbf{Run interruptions and deviations.} The analysis plan, task list and code were frozen before the first M1 run
(15 October 2026), and every amendment is recorded with the plan in the replication package. Harness problems
interrupted M1 three times; each recovery was specified before the run resumed. The first run was discarded before
any outcome was examined, after which output-budget truncation was reclassified as model behaviour. The next two
stopped on a database lock and an HTTP~400 error; their completed tasks form batches 1 and 2, and the remaining 29
tasks ran as batch 3. Tasks unfinished at an interruption, and three tasks per rerun that had finished out of order,
were discarded, and tasks that started within 3{,}600~s of an interruption are flagged as late-started. One first
attempt and one REFLECT retry were lost to provider errors. M2 ran on 29 October without
interruption. M3 ran the same day. Its protocol excluded resumption, but the \texttt{gpt-5.4} part stopped on an HTTP
520 after 16 of 32 tasks, and a first resumption stopped on a local file-read error without completing a task. A
second resumption, which treated HTTP 520--524 as transient, completed the remaining 16 tasks. Both resumptions were specified, reviewed and hash-recorded
before any M3 outcome was examined. The second breaks the first addendum's ``only one resumption'' rule; we chose it
over reporting 44 failures alone. The mini part stopped near its end on a command the sandbox could not execute and
was not resumed (156 of 181 failures). The arms of a failure always ran together, so the paired contrasts do not
span sessions. M3 cost US\$48.8 including discarded work.

\textbf{Scope.} Projects with native extensions and bugs without a linked issue are excluded, so the results apply to
small, pure-Python repositories with issue-style task statements. One project (black) contributes 62 of the 181
failures.

\textbf{Measurement.} The reference fix is one valid solution among others. A retry counts as resolved when the bug-revealing tests pass. The post hoc
full-suite check (Section~\ref{sec:mechanism}) shows that this rate of test-specific patches differs between
conditions; passing the full suite is still not proof of a correct fix~\cite{smith2015cure}, and suites without
network access or with timing tests are incomplete checks. Truncating the failure record is a design
choice that may disadvantage HISTORY and REFLECT on long failures. Six first attempts per task and one retry per
condition limit precision, and more repetitions per task would narrow the intervals only up to a ceiling set by
variation between tasks~\cite{zhou2026repeated}. All results are conditional on a failed first attempt.

\textbf{Generality and contamination.} We studied two models of one family (the replication with only three conditions and on the same
tasks), one harness and one prompt wording per condition, without weight updates. In safety evaluation, scores read at a single canonical wording understate sensitivity to
meaning-preserving rephrasing~\cite{zhou2026single}; by analogy, our estimates may depend on the prompt wording.
BugsInPy fixes are public and may be in the models' training data. Memorised fixes would most likely raise first-attempt
and retry pass rates alike; they could also shrink the room for memory to help, so our estimates may understate its
value on unseen bugs, and we cannot test this with these tasks.

\textbf{Budget.} The harness is tight: eight responses, low reasoning effort and 3{,}000 output tokens per response.
Many attempts ended by using all eight responses (131 first attempts and 79--106 retries per condition,
Table~\ref{tab:subtypes}), so several failures are budget exhaustion rather than wrong fixes, and memory may act
differently under larger budgets. The lessons were written by the same model at the same effort.

\section{Conclusion}
Failure memory did not reliably raise the pass rate of a retry in our study: across 181 failures on 52 of 75 real
bugs, every memory condition had a positive point estimate, but none was significant after correction. The post hoc analyses and follow-ups say more about how
memory changes the retry. On the bug-revealing tests, a model-written lesson showed no advantage over the transcript it
summarises. The transcript exceeded a bare failure notice in point estimate for both models, significantly only on
resolution for the smaller model, in a partial run. Post hoc in M1, fewer retries with memory ended without modifying a Python file (in M2 only clearly for the
transcript), the share of editing retries that touched a fix file was near ceiling in every condition, and resolved
retries in the transcript and digest conditions more often failed other tests that the reference fix passes. Because outcomes cluster strongly by failure, confirmatory studies of effects below about 10 points need larger task sets than
ours. We release the code, frozen inputs and analysis scripts so that the study
can be repeated with other models and larger task sets.

\textbf{Data and code availability.} Code and data are available at
\url{https://github.com/Weypher/agent-failure-memory}: the harness, environment builders, frozen task inputs and prompts, the analysis plan,
per-failure tables and analysis scripts, the post hoc analysis code and rules, the annotation material, and the raw
model requests, responses, tool outputs and test logs of every retained attempt (local paths anonymised). Workspace
snapshots (several GB) will be made available after review.

\section*{Acknowledgment}
The authors used Claude (Anthropic) to assist with code implementation and writing; all content was verified by the authors.

\end{document}